\documentclass[a4paper]{article}

\usepackage{INTERSPEECH2019}
\usepackage[activate]{microtype}
\usepackage{color}
\usepackage{balance}
\usepackage{pifont}
\usepackage{subfig}
\newcommand{\tick}{\ding{51}}

\title{Listen, Attend, Spell and Adapt: Speaker Adapted Sequence-to-Sequence ASR}
\name{Felix Weninger$^1$, Jes\'{u}s Andr\'{e}s-Ferrer$^2$, Xinwei Li$^1$, Puming Zhan$^1$}
\address{
  Nuance Communications, Inc.,
  $^1$Burlington, MA, USA, 
  $^2$Valencia, Spain}
\email{\{felix.weninger,jesusandres.ferrer,xinwei.li,puming.zhan\}@nuance.com}

\begin{document}

\maketitle
\begin{abstract}
Sequence-to-sequence (seq2seq) based ASR systems have shown state-of-the-art performances while having clear advantages in terms of simplicity.
However, comparisons are mostly done on speaker independent (SI) ASR systems, though speaker adapted conventional systems are commonly used in practice for improving robustness to speaker and environment variations. 
In this paper, we apply speaker adaptation to seq2seq models with the goal of matching the performance of conventional ASR adaptation. 
Specifically, we investigate Kullback-Leibler divergence (KLD) as well as Linear Hidden Network (LHN) based adaptation for seq2seq ASR, using different amounts (up to 20 hours) of adaptation data per speaker.
Our SI models are trained on large amounts of dictation data and achieve state-of-the-art results. 
We obtained 25\% relative word error rate (WER) improvement with KLD adaptation of the seq2seq model vs.~18.7\% gain from acoustic model adaptation in the conventional system.
We also show that the WER of the seq2seq model decreases log-linearly with the amount of adaptation data.
Finally, we analyze adaptation based on the minimum WER criterion and adapting the language model (LM) for score fusion with the speaker adapted seq2seq model, which result in further improvements of the seq2seq system performance.
\end{abstract}
\noindent\textbf{Index Terms}: automatic speech recognition, speaker adaptation, sequence-to-sequence, end-to-end

\section{Introduction}

End-to-end ASR systems have recently received increasing attention, due to their ability of integrating all components of an ASR system in a single deep neural network (DNN), which greatly simplifies and unifies the training and decoding process \cite{Graves2012-STW,Battenberg2017-ENT,Li2018-AAT}.
Sequence-to-sequence (seq2seq) modeling \cite{Chan2016-LAA,Gehring2017-CST} is a state-of-the-art method for end-to-end ASR, which has shown competitive results compared to traditional ASR systems \cite{Chiu2018-SOT}. 

However, this kind of modeling technique does not have inherent advantages in terms of robustness to speaker and/or environment changes.
Thus, in this paper, we propose to apply speaker adaptation for seq2seq end-to-end ASR.
Adaptation is commonly applied in practice for conventional ASR, yet it can be 
challenging for end-to-end systems
mainly due to two reasons:
First, because end-to-end ASR subsumes all components in a single DNN, this network usually contains a very large number of free parameters (100\,m and more), which makes it difficult to improve with a small amount of data (say, a few minutes of speech).
Second, since there is no explicit distinction between acoustic model (AM) and language model (LM) in most end-to-end systems\footnote{There are models such as RNNT \cite{Graves:RNNT} that explicitly divide the model into two components which can be understood as equivalent to the LM and AM.}, adaptation can only be done jointly on AM and LM.
This compounds the first problem, as it is generally accepted that effective LM training needs large amounts of text data -- more than might be available in the transcripts of acoustic data used for AM adaptation.

In this regard, our paper makes the following contributions:
To address the first issue, we employ Kullback-Leibler-Divergence (KLD) regularized and Linear Hidden Network (LHN) based adaptation to cope with scarce data.
Moreover, regarding the second issue, we analyze the effects of LM speaker adaptation in combination with shallow fusion~\cite{Chorowsky:Coverage,Kannan:Shallow}.
We show that this adapted LM fusion can effectively exploit additional text data from a speaker in addition to the acoustic adaptation data.
Finally, we demonstrate competitive accuracy compared to a conventional ASR system that exploits both AM and LM adaptation. 

\section{Relation to prior work}

\label{sec:related_work}

Many approaches for adapting `conventional' DNN acoustic models have been developed over the years,
such as linear transformation based approaches in \cite{Neto95-SAF,Gemello2007-LHT,Kumar2015-ILD}, 
training with regularization in \cite{Li2006-RAO,Yu2013-KDR}, 
and various forms of speaker identity vectors in \cite{Saon2013-SAO,Snyder2018-XVR,Vesely2016-SSN}. 
However, relatively few studies so far deal with adaptation of the end-to-end ASR systems.
\cite{Delcroix2018-AFB} proposed to use speaker identity vectors in a hybrid model combining Connectionist Temporal Classification (CTC) with seq2seq modeling.
Furthermore, \cite{Ochiai2018-SAF} investigated the fine-tuning of this model architecture per speaker.
In \cite{Li2018-SAF}, LHN and KLD were investigated for a CTC model.

Regarding LM adaptation, recently there have been several proposals for adapting recurrent neural network (RNN) LMs to domain specific data, including fast marginal adaptation (FMA) \cite{Povey:fma}, as well as noise contrastive estimation (NCE) \cite{Mnih12NCE} combined with KLD regularization~\cite{Ferrer2018-ELM}.
However, these works did not address the interaction with end-to-end ASR, which has an inherent LM adaptation capability.
Note that in \cite{pundak:clas}, a technique similar to LM biasing was proposed for seq2seq model. However, in contrast to our techniques, this does not allow for exploiting large amounts of adaptation data per user.

The novelty of our paper lies in the following aspects:
First, we are not aware of any study applying KLD and LHN adaptation to the seq2seq models with attention.
Furthermore, to our knowledge, there is no previous work comparing the adaptation of end-to-end systems to the offline AM and LM adaptation often used in traditional ASR and demonstrating competitive performance in the speaker adapted scenario. 
Finally, our paper also sheds light on various kinds of LHN adaptation, and introduces the minimum word error rate (mWER) criterion in the context of adaptation as well as adapted LM fusion using KLD + NCE.

\section{Methodology}

\label{sec:methodology}


In this work, we use an encoder-decoder architecture with attention similar to Listen-Attend-Spell (LAS) \cite{Chan2016-LAA}, treating end-to-end ASR as a seq2seq learning task: 
The goal is to predict a sequence $y_i$ of symbols (here, we use sub-word units) from a sequence of acoustic features $x_t$, $t=1,\dots,T$, where $T$ is the number of frames in the utterance.
The encoder $e$ creates a hidden representation $h_j$, $j=1,\dots,T'$ of the acoustic features:
\begin{equation}
    h_1, \dots, h_{T'} = e(x_1, \dots, x_T) .
    \label{eq:encoder}
\end{equation}
In our work, $e$ is implemented as a stack of convolutional (CNN) layers followed by bidirectional Long Short-Term Memory (bLSTM) layers.

The decoder of the seq2seq model is similar to an RNN LM that takes into account a context vector $c_i$ built from the encoder representation for predicting the $i$-th output symbol. 
An attention mechanism is used to focus $c_i$ on various parts of the encoder output sequence.
Here, we use the mechanism proposed by Bahdanau {\em et al.} \cite{Bahdanau2015-NMT}.
The output distribution $p_i = p(y_i | y_1, \dots, y_{i-1}, x)$ for the $i$-th symbol is computed by
\begin{eqnarray}
    s_i &=& f(s_{i-1}, \text{Embedding}(y_{i-1}), c_{i-1}) , \\
    c_i &=& \text{Attention}(s_{i}, h_1, \dots, h_{T'}) , \\    
    s_i' &=& \text{Dense}(s_i, c_i), \\
    p_i &=& g(s_i'), 
    \label{eq:softmax}
\end{eqnarray}
where $g$ is a softmax layer and $f$ is a RNN.
In our work, $f$ is a stack of LSTM layers.
To perform ASR using the seq2seq model, the posterior $p(y|x)$ is factorized as 
\begin{equation}
    p(y|x) = \prod_i p(y_i|y_1,\dots,y_{i-1},x) .
\end{equation}
In practice, beam search is used to produce the sequence $y_1, \dots, y_i, \dots$ of output symbols one at a time.

\subsection{KLD adaptation}

We train speaker adapted (SA) seq2seq models on an adaptation set with transcript $y^*$, initializing with the parameters of the SI model (cf.\ \figurename~\ref{fig:las_adapt}).
The idea of KLD-regularized adaptation \cite{Yu2013-KDR} is to encourage the output distributions of the SA and SI models to be similar, by minimizing the loss:
\begin{equation}
    \mathcal{L}^{\text{KLD}} = \sum_i (1-\beta) \mathcal{L}^{\text{CE}}(y^*_i,p_i) + \beta \mathcal{L}^{\text{CE}}(p_i^{\text{SI}},p_i) .
    \label{eq:kld}
\end{equation}
Here, $\beta$ is the regularization strength (\emph{KLD relevance}), $\mathcal{L}^{\text{CE}}$ is the cross-entropy (CE) loss, and $p^{\text{SI}}_i$ is the output distribution of the SI model.
The target $y^*_i$ is represented as a one-hot vector. 

\subsection{LHN adaptation}

\begin{figure}[t]
    \centering
    \includegraphics[width=.85\columnwidth]{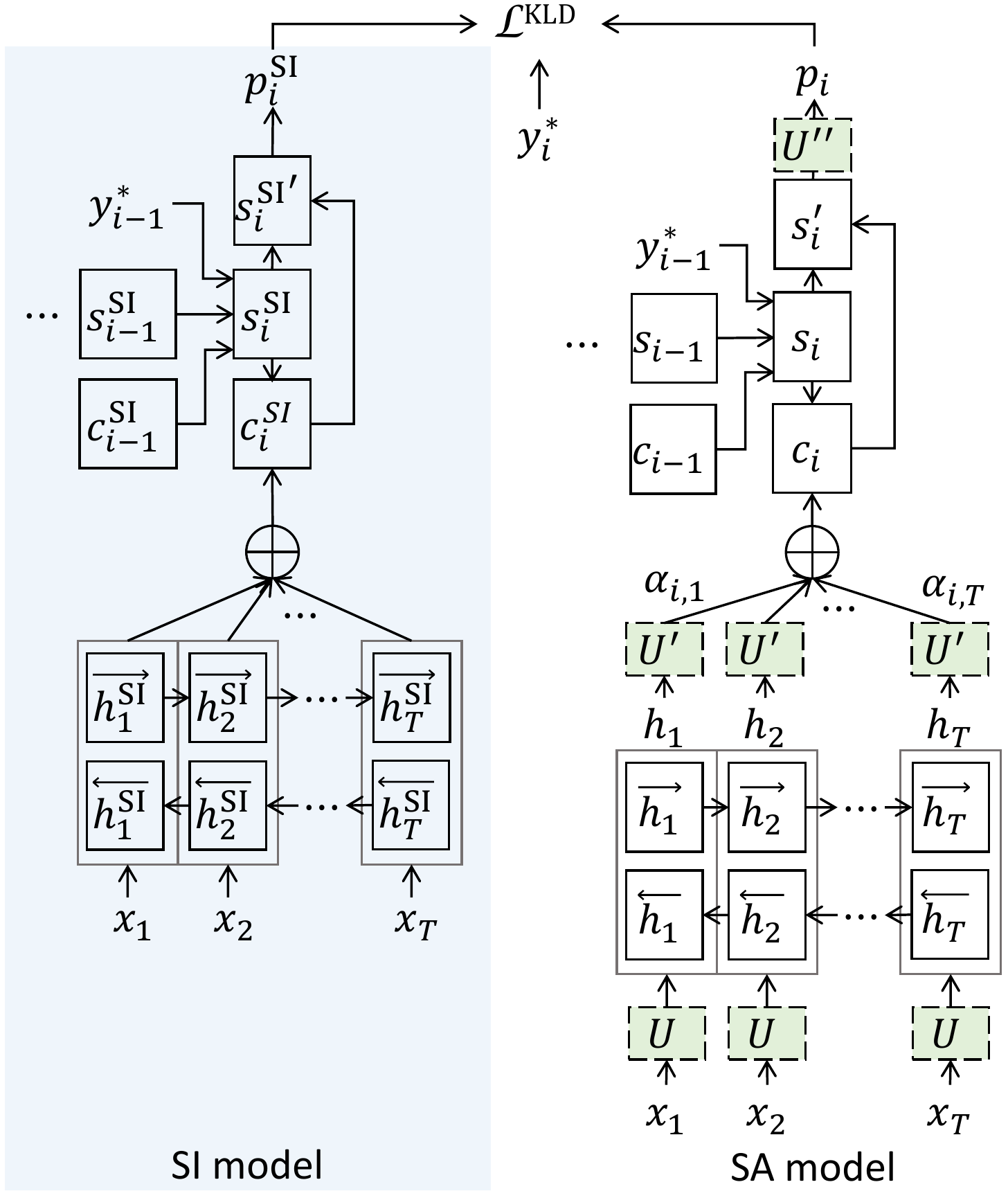}
    \caption{Block diagram of the proposed LAS-like model adaptation via LHN and KLD. 
    For simplicity, a single layer bLSTM encoder is shown instead of the CNN + pyramid bLSTM used in the experiments.
    Dashed boxes indicate possible LHN transformations. 
    $\alpha_{i,\cdot}$: attention weights.
    SI/SA: speaker independent / speaker adapted.}
    \label{fig:las_adapt}
\end{figure}

The idea of LHN adaptation \cite{Gemello2007-LHT,Kumar2015-ILD,Li2018-DOA} is to insert a speaker-specific linear feed-forward layer, which has a square weight matrix $U$.
By initializing $U$ with identity, the starting point for adaptation is the SI model.
During adaptation, only $U$ and the bias $b$ are updated, which reduces the number of speaker-specific parameters compared to KLD adaptation. 
%
In this work, we investigate the application of LHN to various layer outputs in the seq2seq model:
the input features $x$ (cf.~\cite{Gemello2007-LHT}), the encoder output $h_j$, and the decoder output $s'_i$. 
We found it crucial to perform LHN with the KLD regularized loss \eqref{eq:kld}, as shown in \figurename\ \ref{fig:las_adapt}, instead of the simple CE loss.

\subsection{mWER training and adaptation}

mWER training \cite{Prabhavalkar2018-MWE} was introduced as a discriminative training method for seq2seq systems and is similar in spirit to traditional sequence training \cite{Kingsbury2012-SMB}.
%
%
In our experiments, we always build SA models from mWER trained SI models.
Furthermore, in order to exploit mWER training in adaptation, we propose to minimize the KLD loss \eqref{eq:kld} augmented by the mWER loss:
\begin{equation}
    \mathcal{L}^{\text{mWER,KLD}} 
    = \gamma_1 \mathcal{L}^{\text{KLD}} + \gamma_2 \mathcal{L}^{\text{mWER}} .
    \label{eq:mwerkld}
\end{equation}
An alternative way of viewing this is that the mWER loss is regularized with both a CE term (cf.~\cite{Prabhavalkar2018-MWE}; in our case, with weight $\gamma_1 (1-\beta)$) and a KLD term (with weight $\gamma_1 \beta$).

\subsection{LM fusion and decoding}

In order to assess the effect of an external LM in the adaptation of the seq2seq models, 
we use a variation of shallow fusion~\cite{Kannan:Shallow}, which has reported better results than other methods~\cite{Toshniwal:Combination}. 
To avoid the bias of the seq2seq model to both deletions and insertions, we add a coverage score~\cite{Chorowsky:Coverage} per each decoded token  for which the seq2seq models attended more than 0.5 to different encoder  state. 
Specifically, we used beam search for:
\begin{equation}
  \operatorname{arg\,max}_{y}\!\!\big\{ \! \log p(y\!\mid\! x) + \lambda_{\text{LM}} \log p(y) + \lambda_{\text{cov}} \operatorname{cov}(x,y) \big\}
\end{equation}
where $p(y | x)$ is the seq2seq probability, $p(y)$ is the LM probability,  $\operatorname{cov}(x,y)$ is the coverage score in Eq.~(11) of~\cite{Chorowsky:Coverage},
and $\lambda_{\text{LM}}$ and $\lambda_{\text{cov}}$ are the weights for the LM and the coverage score.

\section{Experiments}

\label{sec:experiments}

\subsection{Data set}

We perform our experiments on a dictation data set.
All utterances are anonymized field data.
The audio is sampled at 8\,kHz.
A training set of 7.6\,k hours from 58\,k speakers is used to train SI models. 
The performance is measured on an evaluation set with 35 speakers (392\,k words).
The speakers cover various dictation domains with different vocabularies.

For each speaker in the evaluation set, we have up to 20 hours of acoustic adaptation data available.
We also perform experiments where the adaptation data amount is restricted to 5 minutes, 30 minutes, 2 hours, or 6 hours.
Besides, for LM adaptation, additional text documents (0.37\,m on average per speaker) related to the dictations are available.


The transcripts of the training set as well as the adaptation sets are pseudo truth, i.e., the output of an ASR system along with some user corrections.
Thus, the adaptation schema is semi-supervised.
In contrast, the evaluation set is manually transcribed.

\subsection{End-to-end models}

Our ASR systems use 40-dimensional log Mel filterbank outputs as input features.
In the seq2seq model, the inputs are first passed through a stack of 3 CNN layers, which is parameterized so as to yield a 768-dimensional embedding for each input frame.
Subsequently, there is a pyramid bLSTM encoder with 6 layers (768 LSTM units per layer and direction), which performs frame decimation (by a factor of 2) after every other layer, thus reducing the frame rate by a factor of 8.
The decoder uses 2 (unidirectional) LSTM layers (1\,536 units per layer).
The softmax output layer predicts the posterior probabilities for 20\,k word piece targets determined on the training set. 
In total, the model has 181\,m parameters.

The SI end-to-end model is trained 
using the CE criterion, followed by mWER training \cite{Prabhavalkar2018-MWE}.
We employ dropout \cite{Srivastava2014-DAS} (with probability 0.1), label smoothing \cite{Szegedy2016-RTI} (with probability 0.9 for the target class), and early stopping (using a validation set held out from the training data) in order to improve generalization of the model.

For the adaptation experiments, the KLD relevance is set to $\beta=0.6$ and a small fixed number of epochs is run. 
In adaptation, we decrease the learning rate to perform fine-tuning, increase the dropout probability to 0.25 to cope with the limited amount of data, and 
we decrease the batch size, so that the model is updated more frequently.
The adaptation hyperparameters were tuned in preliminary experiments separately for the end-to-end and the baseline models. 

For LM fusion, we trained a LSTM LM with embeddings of 512 and 2 LSTM layers of 1\,k units each, followed by a linear projection to 512. 
The outputs of the LM correspond to the word pieces of the seq2seq model. 
The LM is self-normalized and trained with the NCE~\cite{Mnih12NCE} loss. 
The LSTM LM is trained by sampling a large training set comprising 33.4 billion running words.
We regularized the model applying dropout to the feed-forward connections~\cite{zaremba2014:ffnn} and some small ($10^{-6}$) weight decay. 
For adapting the LM per speaker, we use the transcripts of the 20\,h acoustic data and optionally the speakers' text documents to fine-tune the LM with a small learning rate, and optionally also regularize the NCE loss by adding a KLD term~\cite{Ferrer2018-ELM}.

\subsection{Baseline models}

We compare the performance of the end-to-end system against a `conventional' hybrid ASR system, which uses a linearly augmented (LA-) DNN AM \cite{Ghahremani2016-LAD} with 22 layers and 25\,m parameters.
The acoustic features are the same as in the end-to-end system.
The SI LA-DNN AM is initially trained using the CE loss, followed by Hessian-free sequence discriminative training \cite{Kingsbury2012-SMB}.
Then, we train SA LA-DNN AMs using the KLD-regularized CE loss in analogy to 
\eqref{eq:kld}, with $\beta=0.2$.

We use a specific LM for each of the dictation domains in the evaluation set.
Each LM comprises n-gram as well as RNN components and is trained on several hundred millions of words, collectively summing up to billions of words.
The domain specific LMs are then interpolated with speaker specific LMs, which are trained on the same data that is used to fine-tune the LSTM LM used in seq2seq LM fusion.

\section{Results}

\label{sec:results}

\begin{table}[t]
    \caption{WER [\%] and WERR [\%] compared to the SI model with the KLD method, adapting all parameters of the seq2seq model or subsets thereof with 2\,h of adaptation data.}
    \label{tab:adaptation_parts}
    \centering
    \begin{tabular}{c|c|cc}
       Parameters & \# Parameters & WER & WERR \\
       \hline
       Encoder ($e$) & 83.0\,m & 9.60 & 7.7 \\
       Decoder ($a, f, g$) & 98.0\,m & 9.12 & 12.3 \\ 
       All & 181\,m & \bf 8.91 & \bf 14.3 \\
    \end{tabular}
\end{table}

\begin{table}[t]
    \caption{WER and WERR compared to the SI model with the LHN method for various positions of the linear layer in the seq2seq model, adapting with 2\,h of data.}
    \label{tab:adaptation_lhn_pos}
    \centering
    \begin{tabular}{c|c|cc}
       LHN applied to & LHN dim. & WER & WERR \\
       \hline
       Features ($x$) & 40$\times$40     & 10.31    & 0.9  \\
       Encoder ($h$)  & 1536$\times$1536 & 9.56     & 8.1 \\
       Decoder ($s'$) & 1536$\times$1536 & \bf 9.23 & \bf 11.3 \\
    \end{tabular}
\end{table}

\subsection{Adaptation of various parameter subsets}

We start our evaluation by adapting different parameter subsets of the seq2seq model with the KLD method, using 2\,h of adaptation data.
We conjecture that due to the importance of language modeling for the dictation task, adapting only the {\em decoder} should yield a significant gain as well.
As can be seen from \tablename\ \ref{tab:adaptation_parts}, adapting only the decoder increases the WER by only about 2\,\% relative compared to adapting the entire model.
At the same time, it is much less expensive, since the error signal does not have to be backpropagated through the deep bLSTM encoder stack.
Conversely, adapting only the encoder as in \cite{Ochiai2018-SAF} is not competitive on our data set. 

\subsection{Position of the LHN layer}

Next, Table \ref{tab:adaptation_lhn_pos} shows the performance of the LHN method, varying the position of the inserted linear layer. 
We found that inserting the LHN transformation after the acoustic features has limited potential. 
Using LHN as encoder output transformation performs slightly better than KLD adaptation of the encoder parameters.
Moreover, using LHN as decoder output transformation delivers the most promising results, yielding 11.3\,\% relative WER reduction (WERR).
It is similar in performance to decoder KLD adaptation (see \tablename\ \ref{tab:adaptation_parts}),
although only a small fraction of the parameters are adapted (2.4\,m vs.\ 98\,m parameters).

\subsection{Influence of the amount of adaptation data}

\begin{figure}
    \centering
    \includegraphics[width=\columnwidth]{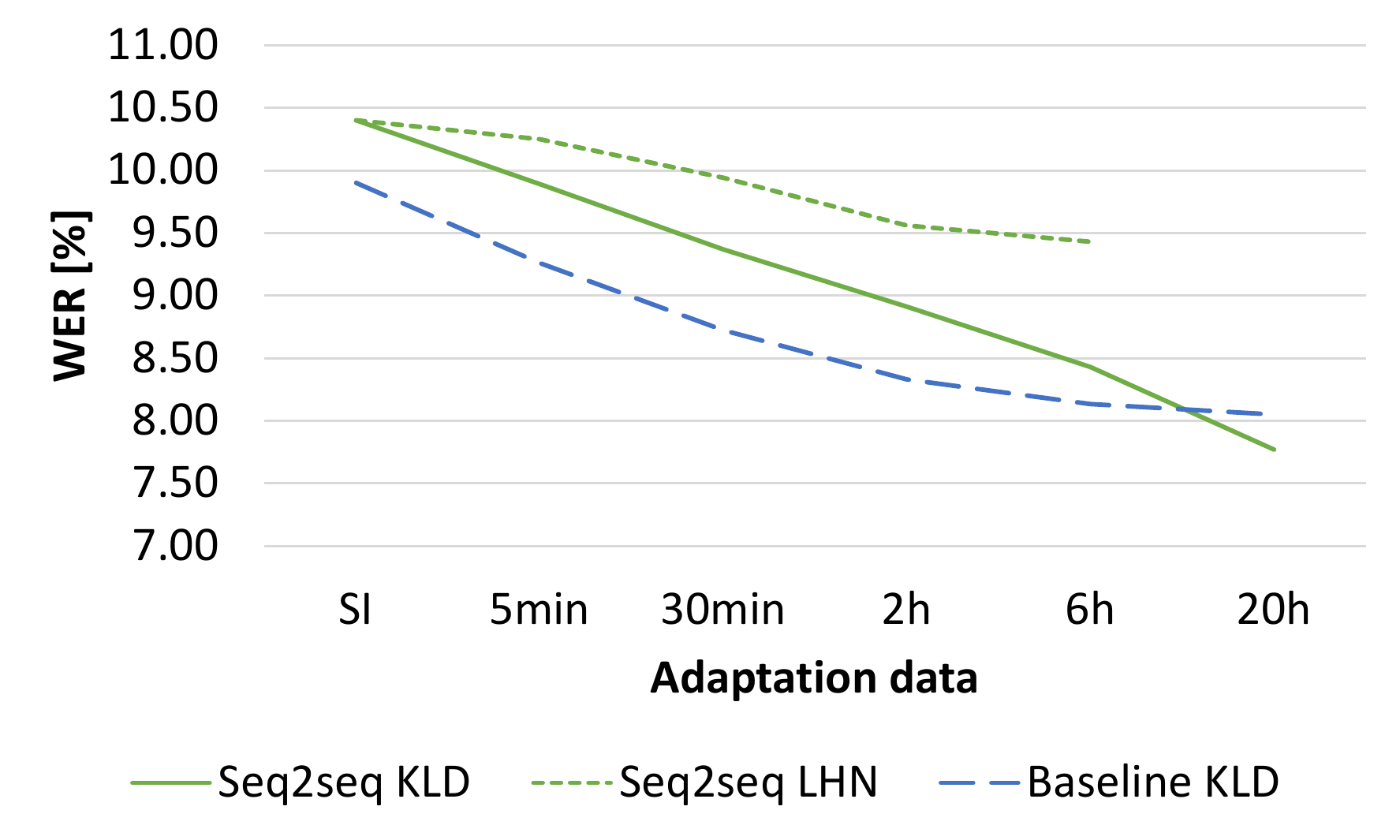}
    \caption{WER of the speaker adapted seq2seq system (KLD or encoder LHN adaptation) and the speaker adapted conventional ASR system (KLD) with various amounts of adaptation data.}
    \label{fig:adaptation_data}
\end{figure}

We now compare the results with KLD and LHN for various amounts of adaptation data.
As can be seen from \figurename\ \ref{fig:adaptation_data}, the WER decreases log-linearly in the amount of adaptation data with KLD adaptation, yielding 7.77\,\% WER (25.3\,\% WERR) with 20\,h of data.
Regarding LHN, one could conjecture that it could be more effective than KLD for little adaptation data, as it adapts less parameters.
For 5 minutes data, we found encoder LHN to be more stable than decoder LHN.
Contrary to our expectations, LHN was outperformed by KLD on all the data points shown in Figure \ref{fig:adaptation_data}, 
showing 
the effectiveness of the KLD regularization.

We also compare the performance to KLD regularized AM adaptation of the conventional system with various amounts of data.
We can observe that the performance starts saturating when adapting with more than 2\,h of data; as a result, the SA seq2seq system outperforms the SA conventional ASR system with 20\,h of adaptation data.
We attribute this to implicit LM adaptation in the seq2seq system. 
Thus, we will also compare the seq2seq and conventional systems under LM adaptation in Section \ref{sec:lm_fusion}.

\subsection{Minimum WER adaptation}

Further, \tablename\ \ref{tab:mwer} shows the results obtained by using the KLD + mWER criterion \eqref{eq:mwerkld} in adaptation (using $\gamma_1 = \gamma_2 = 1$), starting from KLD SA models. 
We found some inconsistencies in the formatting of the adaptation vs.\ the evaluation data, to which the mWER adaptation was particularly sensitive.
Using the cleaned data, the standard KLD criterion achieves 8.70\,\% WER with 2\,h data, while we obtain 8.10\,\% WER from the mWER + KLD criterion.
With stronger regularization ($\beta=0.8$), the WER is slightly higher, but the improvement is more uniform (no degradations for any speaker). 
Increasing the adaptation data to 20\,h, we obtain 7.33\,\% WER, which is a 3.2\,\% relative improvement compared to KLD (7.57\,\% WER).

\begin{table}[t]
    \caption{WER [\%] and WERR [\%] compared to the SI model using mWER + KLD \eqref{eq:mwerkld} by amount of data and regularization strength ($\beta$).} 
    \label{tab:mwer}
    \centering
    \begin{tabular}{c|c|cc}
      \# Data & $\beta$ & WER & WERR \\
      \hline
      2\,h   & 0.6 & 8.10 & 22.1 \\
      2\,h   & 0.8 & 8.23 & 20.9 \\
      20\,h  & 0.8 & \bf 7.33 & \bf 29.5 \\
    \end{tabular}
    \vspace{-3mm}
\end{table}

\subsection{Fusion with adapted LM}

\label{sec:lm_fusion}

To verify the gains from LM fusion and adaptation in combination with seq2seq model adaptation, which already adapts the decoder part, we combined a seq2seq model adapted on 2\,h acoustic data (using decoder  LHN, cf.\ \tablename\ \ref{tab:adaptation_lhn_pos}) with the generic LSTM LM and compared to the adapted LSTM LM.
The results are shown in \tablename\ \ref{tab:lm_fusion}.
We were able to obtain 3.9\,\% WERR from LM fusion and 2.4\,\% WERR from LM adaptation on top of LM fusion.
The latter shows that LM adaptation can help incorporating additional user-specific data into the seq2seq model.
Moreover, there is an additional 2.1\,\% WERR from using the NCE + KLD LM adaptation instead of simple fine-tuning.


\tablename\ \ref{tab:overview} shows the summary of results achieved by conventional and seq2seq ASR with and without speaker adaptation, now using 20\,h of data.
While the seq2seq model without external LM is behind the conventional system, fusion with the generic LSTM LM makes it competitive with the conventional baseline (9.75 vs. 9.90\,\% WER).
Applying LM fusion to the KLD + mWER adapted seq2seq system, we obtain 7.03\,\% WER (4.1\,\% WERR), which is competitive with the conventional ASR using AM and LM adaptation.
While adaptation of the LM used for LM fusion improves significantly upon the SI seq2seq system (9.15 vs.\ 9.75\,\% WER), and LM adaptation yields a considerable improvement upon the conventional system with adapted AM (7.18 vs.\ 8.04\,\% WER), there is only a small gain for the SA seq2seq system (6.95 vs.\ 7.03\,\% WER).
However, this gain is remarkable because the decoder-internal LM adaptation on 20\,h transcripts is already done in the seq2seq adaptation. 

\begin{table}[t]
    \caption{WER [\%] and WERR [\%] compared to seq2seq model adaptation only (decoder LHN, 2\,h data) by LM fusion and LM adaptation (20\,h adaptation data transcripts and user text).}
    \label{tab:lm_fusion}
    \centering
    \begin{tabular}{l|c|c}
    Shallow fusion LM & WER      & WERR \\
    \hline
    None & 9.23 & -- \\\hline
    LSTM 33\,b words & 8.87 & 3.9 \\
    \;\;+ LM adaptation &  8.66 &  6.2 \\
    \;\;+ KLD & \bf 8.48 & \bf 8.1 \\
    \end{tabular}
    \vspace{-3mm}
\end{table}

\section{Conclusions}

\label{sec:conclusions}

In this paper, we have presented several effective techniques for adapting seq2seq ASR systems.
Significant gains have been achieved even with a few minutes of speech data, and at the same time we have shown that seq2seq ASR systems can exploit larger amounts of adaptation data effectively.
Furthermore, 
we have achieved state-of-the-art performances in the SA scenario, comparing to conventional ASR.
Nevertheless, we also found that it is hard to get complementary gains from LM adaptation on top of seq2seq adaptation. 
To improve on this, we will look into 
interpolation of generic and and speaker adapted LMs.
Moreover, in future work,
we plan to explore 
LHN and i-vector on-line adaptation similar to \cite{Li2018-DOA}.

\begin{table}[t]
    \caption{Comparison of conventional and seq2seq ASR with and without speaker adaptation, using 20\,h of acoustic data for seq2seq / AM adaptation (via mWER + KLD) and 20\,h transcripts + user documents for LM adaptation.}
    \label{tab:overview}
    \centering
    \subfloat[Baseline system]{
    \begin{tabular}{cc|c}
        \multicolumn{2}{c|}{Adaptation} & WER \\
        AM & LM & \\
        \hline
        -- &  -- & 9.90 \\
        \tick & -- & 8.04 \\
        -- & \tick & 7.97 \\
        \tick & \tick & \bf 7.18 \\
    \end{tabular}
    }
    \subfloat[seq2seq system]{
    \begin{tabular}{c|cc|c}
        LM & \multicolumn{2}{c|}{Adaptation} & WER \\
        fusion & seq2seq & LM & \\
        \hline
        -- & -- &  -- & 10.40 \\
        -- & \tick & -- & 7.33 \\
        \tick & -- & -- & 9.75 \\
        \tick & \tick & -- & 7.03 \\
        \tick & -- & \tick & 9.15 \\
        \tick & \tick & \tick & \bf 6.95 \\  
    \end{tabular}
    }
    \vspace{-3mm}
\end{table}

\section{Acknowledgements}

We would like to thank Peter Skala and Ming Yang for their help with the baseline ASR adaptation and many helpful discussions.


\bibliographystyle{IEEEtran}

\balance

\bibliography{mybib}

\end{document}